\documentclass[preprintnumbers]{revtex4}
\UseRawInputEncoding
\usepackage{amssymb}
\usepackage{amsmath}
\usepackage{graphicx}
\usepackage{dcolumn}
\usepackage{bm}
\usepackage{subfigure}
\usepackage{color}

\begin{document}

\title{Overview of thermodynamic properties for Reissner-Nordstrom-de Sitter spacetime in induced phase space}
\author{Yun-Zhi Du$^{1,2}$, Huai-Fan Li$^{1,2}$\footnote{the corresponding author}, Ren Zhao,$^{2}$}
\address{$^1$Department of Physics, Shanxi Datong University, Datong 037009, China\\
$^2$Institute of Theoretical Physics, Shanxi Datong University, Datong, 037009, China }

\thanks{\emph{e-mail:duyzh22@sxdtdx.edu.cn, huaifan999@163.com, zhao2969@sina.com}}

\begin{abstract}
Since the black hole and cosmological horizons in the de Sitter (dS) spacetime with the Reissner-Nordstrom (RN) black hole are not independent to each other, which is caused by the gravitational effect, the interplay between two horizons should be considered. Based on this, by introducing the interactive entropy the RN-dS spacetime is analogous to a thermodynamic system with various thermodynamic quantities, in which the thermodynamic laws still hold on. In our work, the thermodynamic properties of the RN-dS spacetime are mapped out in the induced phase space, which are similar to that in AdS black holes. The phase transition of the RN-dS spacetime between the high-potential and the low-potential black hole phases is observed. Compared with an ordinary thermodynamic system, the similar behaviours about the Joule-Thomson expansion and the critical exponents are also checked out. Finally, the scalar curvature of two existent phases are presented to reveal the underlying microstructure and nature of phase transition in the RN-dS spacetime, which opens a new window to investigate the dS spacetime with black holes from a observational perspective.
\end{abstract}

\maketitle

\section{Introduction}
With the development of theory, people found there is a fundamental relationship in gravity, thermodynamics and quantum theory. Initially, people studied the mechanics of black holes \cite{Bardeen1973}, and they found that if applying the quantum field theory in curved spacetime to black holes, they could get the profound results. The black hole thermodynamics is based on the discovery made by Bekenstein and Hawking that black holes have entropy and other thermodynamic quantities \cite{Bekenstein1972,Bekenstein1973a,Hawking1975,Bekenstein1974}. The black holes have negative specific heat in asymptotically flat spacetime which causes it to be an unstable system. The problem in asymptotically flat space then can be avoided since the gravitational potential of dS/AdS spacetime acts as a box of finite volume with unphysical perfectly reflecting walls \cite{Hawking1983}. The AdS-Schwarzschild black hole was investigated by Hawking and Page, where the phase transition from thermal AdS to black hole was found. By introducing the AdS/CFT correspondence \cite{Witten1998} or the AdS/QCD correspondence \cite{Pourhassan2013,Sadeghi2013}, the phase structure of AdS black holes becomes more charming. Motivated by that fact, and the recent observations of the striking similarity between the thermodynamic phase structure of AdS black holes in the canonical ensemble and that of the van der Waals-Maxwell liquid-gas system, the physics of the charged AdS black holes were explored in more detail \cite{Chamblin1999, Chamblin1999a}. Regarding the negative cosmological constant as the thermodynamic pressure \cite{Kastor2009,Dolan2011,Cvetic2011} enriches the phase structure of black holes \cite{Kubiznak2012,Cai1306,Wei1209,Hendi1702,Bhattacharya2017,Zhang1502,Zou1702}.
In line with that, the phase transition and critical behavior of the black hole in the extended phase space have been numerously investigated in \cite{Toledo2019,Hendi2013,Wei2013,Cai2013,Zhao2013,Mo2013,Altamirano2013,Spallucci2013,Xu2014,Miao2018,Miao2017,Xu2014a}.

In another hand, there is a correspondence between a strong gravitational theory in de Sitter space and conformal field theory \cite{Strominger2001}. The thermodynamics study of a dS spacetime with black hole is interesting and important see Ref. \cite{Torii1999}. That is due to the fact that the universe during the early-inflationary epoch was a dS spacetime, in far future it will turn into a dS one \cite{Weinberg2008,Mukhanov2005}. Furthermore, similar to the AdS/CFT correspondence in AdS spacetime with a black hole, the dS/CFT correspondence was apply to dS spacetime with a black hole in Ref. \cite{Kubiznak2016}, in which more information about possible phase transitions in various dimensions was explored. Under these efforts, black holes in dS/AdS spacetime have been widely recognized as thermodynamic systems. What's more, the observation of black hole opens a new window for its thermodynamic research.

While black holes in dS spacetime, in which the effect of black hole inner horizon is often neglected, cannot be in thermodynamic equilibrium in general due to the multiple horizons with different temperatures. To overcome this problem, one had analyzed one horizon and taken another one as the boundary or separated the two horizons by a thermally opaque membrane or box \cite{Gomberoff2003,Sekiwa2006}, in which the two horizons were analyzed independently. Besides, one could take a global view to construct the globally effective temperature and other effective thermodynamic quantities \cite{Urano2009,Zhao2014}. In this issue, there are three assumptions of constructing the total entropy in dS spactime with black hole: i) the sum of both horizons \cite{Kastor1993,Bhattacharya2016} with the volume $V_{eff}=V_c-V_b$. In this case the black hole and cosmology horizons are independent with each other; ii) the difference of both horizons \cite{Kubiznak2017,Kanti2017} with $V_{eff}=V_c+V_b+V_{in}$; iii) the sum of both horizons and their interaction with $V_{eff}=V_c-V_b$ \cite{Zhang2016,Ma2018,Zhao2021}. The revelent works on the thermodynamic property of dS spacetime with black hole were presented \cite{Dolan2013,Cai2002,Cai2002a,Carlip2003,Braden1990,Simovic2018,Mbarek2019,Guo2020a,Zhang2020,Ma2020,Dinsmore2020}. Note that due to the existence of gravity in system, the interplay between the black hole and cosmology horizons should be considered, thus we adopt the third assumption in this work. In the other hand, for the space between the black hole outer horizon (generally called as the black hole horizon) and cosmology horizon, the different hawking temperatures on two horizons prevents a dS spacetime with black hole in equilibrium as like an ordinary thermodynamic system. While, there are common parameters $M$, $Q$, $¦«$ on the black hole and cosmological horizons. Thus, the thermodynamic quantities on two horizons are not independent. The interplay between two horizons must be taken into account when constructing the effective quantities of a dS spacetime with black hole in thermodynamic equilibrium. Based on these, the thermodynamic phase transition, Joule-Thomson expansion, and the continuous phase transition theory for the RN-dS spacetime will be further probed.

In addition, despite the success of black hole thermodynamics, its microscopic origin is a mystery at present. The Ruppeiner geometry \cite{Ruppeiner1981} derived from the laws of thermodynamic fluctuations gives us a glimpse of the microstructure of black holes. The Ruppeiner line element measures the distance between two adjacent thermodynamic fluctuation states, from which we can obtain the thermodynamic Riemannian curvature scalar. The corresponding curvature scalar marks the interaction between adjacent modules of fluid systems: a positive (negative) thermodynamic curvature scalar indicates that a repulsive (attractive) interaction domain and a noninteracting system such as the ideal gas corresponds to the flat Ruppeiner metric \cite{Ruppeiner2008,Sahay2010,Wei2017,Chaturvedi2018,Xu2020}. Thus, in this work, we will investigate microstructure of the RN-dS spacetime near the phase transition point by the Ruppeiner geometry.

The work is organized as follows. In Sec. \ref{two}, we would like to discuss the conditions for the existence of black hole and cosmological horizons in RN-dS spacetime, as well as the influence of spacetime parameters, and give the range of the position ratio $x$ of two horizons. Then present the effective thermodynamic quantities and the critical point of the RN-dS spacetime, analyze the behaviors of the heat capacities, isobaric expansion coefficient, and isothermal compression coefficient. In Sec. \ref{three}, based on the effective thermodynamic quantities of RN-dS spacetime, we extend the method of the phase transition for Van de Waals systems or charged AdS black holes, use it to study the RN-dS spacetime, and exhibit the first-order phase transition of the RN-dSQ spacetime. In Sec. \ref{four}, we explore the Joule-Thomson expansion of the RN-dS spacetime in isenthalphy processes. The inversion curves divide the isenthalpic ones into two parts in the $P-T$ diagram: a cooling phenomena with the positive slope and a heating process with negative slope. In Sec. \ref{five}, the critical exponents of the RN-dS spacetime are obtained and they are satisfied with the universal relations. In Sec. \ref{Six}, we investigate the thermodynamic geometry of two coexistent phases for the RN-dS spacetime. The result shows the geometries of two coexistent phases are both positive, and the geometry of one phase is always bigger than the other phase. That means the RN-dS spacetime in two coexistent phases are domain with the repulsive interaction, and the average interaction of this system with one phase is bigger than the other phase. This phenomenon provides a new way for us to further study the microstructure interaction of dS spacetimes. The Sec. \ref{Seven} is the summary.

\section{Review of Stable RN-dS Spacetime}
\label{two}
We will review the solution of the RN black hole embedded in a dS spacetime. And we give the concrete process of exporting the effective thermodynamic quantities and the critical point of RN-dS spacetime in the induced phase space, and analyze the heat capacities, isobaric expansion coefficient, and isothermal compression coefficient, respectively.

\subsection{Solution of RN black hole in dS spacetime}
\label{2.1}
The static spherically symmetric solution of Einstein equation for a RN-dS black hole have been obtained
\begin{equation}
d s^{2}=-f(r) d t^{2}+f^{-1} d r^{2}+r^{2} d \Omega_{2}^{2}
\end{equation}
with the horizon function
\begin{equation}
f(r)=1-\frac{2 \widetilde{M}}{r}+\frac{Q^{2}}{r^{2}}-\frac{r^{2}}{l^{2}},\label{fr}
\end{equation}
where $\widetilde{M}$ and $Q$ are the black hole mass and charge, $l$ is the curvature radius of dS spacetime. There are three horizons: the RN-dS black hole inner horizon, the RN-dS black hole outer horizon (i.e., RN-dS black hole horizon), and the cosmological horizon. The radiuses of RN-dS black hole horizon and the cosmological horizon satisfy the expression $f(r_+,r_c)=0$. The mass parameter can be obtained from the above equation
\begin{eqnarray}
\widetilde{M}&\equiv&QM=\frac{r}{2}\left(1+\frac{Q^2}{r^2}-\frac{r^2}{l^2}\right),
\end{eqnarray}
where $M$ is the reduced mass of black hole.

\begin{figure}[htp]
\subfigure[$M-r$]{\includegraphics[width=0.4\textwidth]{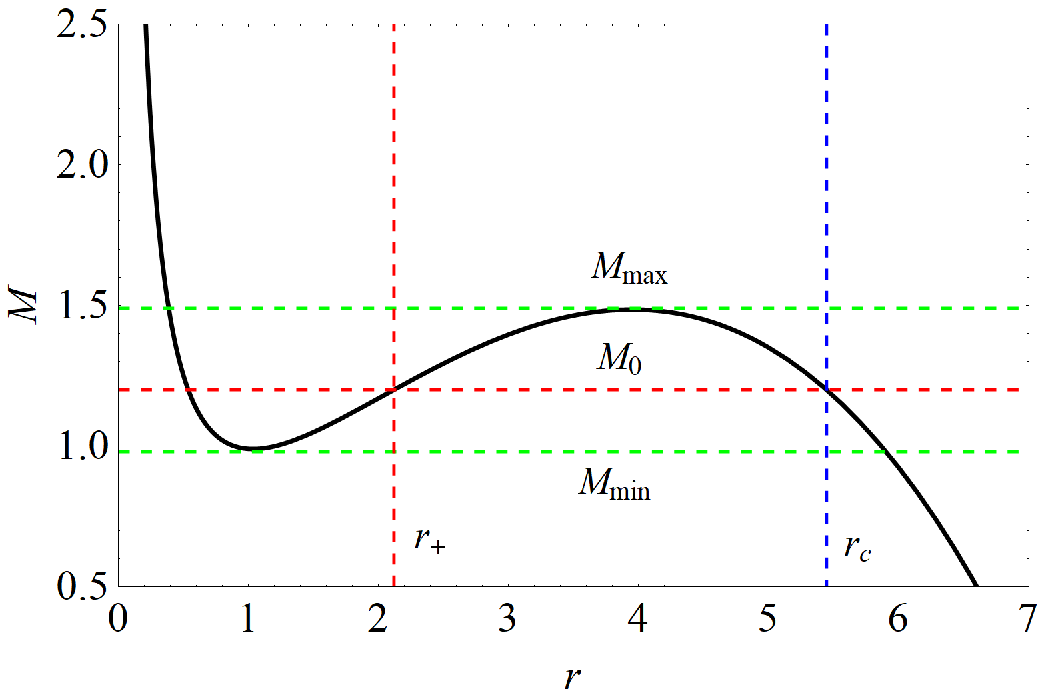}\label{Mr}}~~~
\subfigure[$M-\phi$]{\includegraphics[width=0.4\textwidth]{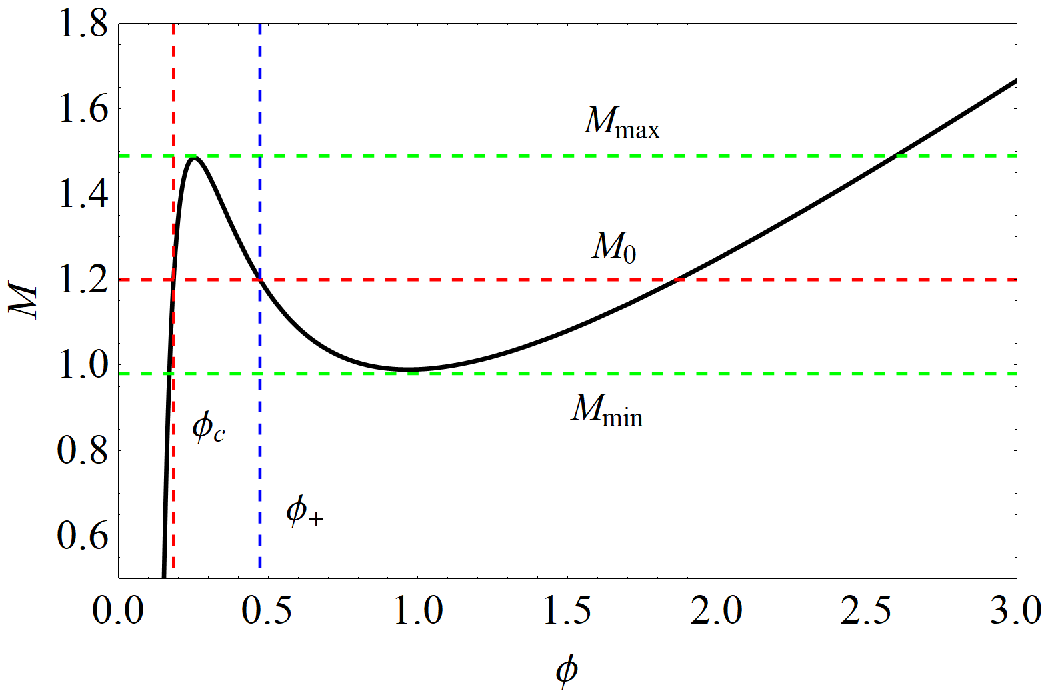}\label{Mphi}}
\caption{The pictures of $M-r$ and $M-\phi$ with the parameter $\frac{Q^2}{l^2}=\frac{1}{50}$.}\label{Mrphi}
\end{figure}

For similarity, we introduce the electric potential $\phi=Q/r$, the metric function of (\ref{fr}) can be rewritten as
\begin{equation}
f(r)=1-2 M\phi +\phi^{2}-\frac{Q^{2}}{l^{2}\phi^{2}}.\label{fphi}
\end{equation}
On two horizons, the potentials read $\phi_+=Q/r_+$ and $\phi_c=Q/r_c$. The reduced mass parameter becomes
\begin{eqnarray}
M=\frac{1}{2\phi}\left(1+\phi^2-\frac{Q^2}{l^2\phi^2}\right).\label{M}
\end{eqnarray}
The plots of $M-r$ and $M-\phi$ with the parameter $\frac{Q^2}{l^2}=\frac{1}{50}$ are given in Fig. \ref{Mrphi}.

\begin{figure}[htp]
{\includegraphics[width=0.4\textwidth]{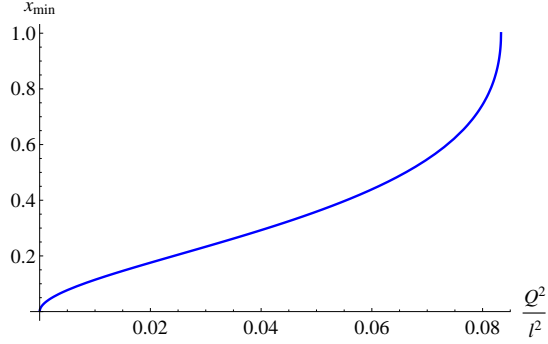}}
\caption{The picture of $x_{min}-\frac{Q^2}{l^2}$.}\label{xmin}
\end{figure}

For the RN-dS spacetime, there exist the local minimum ($M_{min}$) and local maximum ($M_{max}$) for black hole mass. As $M_{min}\leq M_0\leq M_{max}$, this system have three horizons, i.e., the black hole inner and outer horizons, the cosmological horizon. Note that the black hole horizon $r_+$ means the outer one. When the local minimum and local maximum merge into an inflexion point, these three horizons coincide together. In this case, the hole black is called the ultracold black hole ($r_{ucold}=\sqrt{2}Q,~\frac{Q^{2}}{l^{2}}=\frac{1}{12},~M_{ucold}=\frac{2\sqrt{2}Q}{3}$). Generally, the RN-dS black hole horizons do not coincide with the cosmological one, for a RN-dS system the conditions $\frac{Q^{2}}{l^{2}}\leq\frac{1}{12}$ and $M_{min}\geq M_{ucold}$ should be held on. In the following we will investigate the thermodynamic property for this system from the view of the potential on horizon, instead of horizon radius. From eq. (\ref{M}) and the definition $x\equiv\frac{\phi_c}{\phi_+}$, we have
\begin{eqnarray}
\frac{Q^2}{l^2}=\frac{x^2}{1+x+x^2}\phi^2_+-\frac{x^3}{1+x+x^2}\phi^4_+.\label{ql2}
\end{eqnarray}
When $M_0=M_{min}>M_{ucold}$: the black hole inner and outer horizons coincide together, such black hole is called the cold one. As $M_0=M_{max}>M_{ucold}$: the black hole and cosmologic horizons coincide together ($x=1$), this black hole is called the Nariai one whose maximum potential becomes
\begin{eqnarray}
\phi_{+max}^{2}=\frac{1}{2}\left(1+\sqrt{1-\frac{12Q^2}{l^2}}\right).\label{phimax}
\end{eqnarray}
Generally, the potential on horizon is a limited value, not infinity, i.e., the ratio of two horizon potentials should be from a minimum instead of zero to one. For the Nariai black hole, substituting eq. (\ref{phimax}) into eq. (\ref{ql2}), the minimum of $x$ is constrained by the following expression
\begin{eqnarray}
\frac{x_{min}^2}{1+2x_{min}+3x_{min}^2}=\frac{1}{6}\left(1-\sqrt{1-\frac{12Q^2}{l^2}}\right)\label{x}
\end{eqnarray}
with $\frac{Q^2}{l^2}\leq\frac{1}{12}$, whose picture is shown in Fig. \ref{xmin}. In this system, the RN-dS black hole can be survives with the range of $x_{min}\leq x\leq1$.

\subsection{Effective thermodynamic quantities of RN-dS Spacetime}
\label{2.2}
For a dS spactime with RN black hole, one can describe the thermodynamical property on the black hole and cosmological horizons, respectively, where the thermodynamical laws are still satisfied. Since the first law are held on for the black hole and cosmological horizons, the thermodynamic quantities on two horizons are
\begin{eqnarray}
\widetilde{T}_c&\equiv&\frac{T_c}{Q}=-\frac{x\phi_+}{4\pi Q}\left(1-x^2\phi_+^2-\frac{3Q^2}{x^2l^2\phi_+^2}\right),
~\widetilde{S}_c\equiv Q^2S_c=\frac{Q^2\pi}{x^2\phi_+^2},~~~\widetilde{V}_c\equiv Q^3 V_c=Q^3\frac{4\pi}{3 x^3\phi_+^3},\\
\widetilde{T}_+&\equiv&\frac{T_+}{Q}=\frac{\phi_+}{4\pi Q}\left(1-\phi_+^2-\frac{3Q^2}{l^2\phi_+^2}\right),~~~~~~~
~\widetilde{S}_+\equiv Q^2S_+=\frac{Q^2\pi}{\phi_+^2},~~~\widetilde{V}_+\equiv Q^3 V_+=Q^3\frac{4\pi}{3\phi_+^3}.
\end{eqnarray}
When the hawking radiation temperatures on two horizons are equal to each other, i.e., in the lukewarm case \cite{Torii1999} the potentials on two horizons become
\begin{eqnarray}
\phi_+^2=\frac{1}{(1+x)^2},~~~~\phi_c^2=\frac{x^2}{(1+x)^2},
\end{eqnarray}
and the same radiation temperature on two horizons reads
\begin{eqnarray}
\widetilde{T}&=&\widetilde{T}_c=\widetilde{T}_+=\frac{2x(1-x)}{4\pi Q(1+x)^3}, ~~~or ~~~T=T_c=T_+=\frac{2x(1-x)}{4\pi (1+x)^3}.
\end{eqnarray}

Since there are two different Harking-temperature on the RN black hole and cosmological horizons, we can not directly regard the space between the black hole and cosmological horizons as an ordinary thermodynamic system in equilibrium to study the phase transition. While considering the gravity effect between two horizons' space, i.e., the interplay of two horizons should be introduced, the RN-dS spacetime with the space between two horizons can be treated as a thermodynamic system affected by gravitational effects, where the thermodynamical laws are still held on. From the view of the whole RN-dS spacetime, two horizons are not independent with each other, because they are in gravity field which will leads to the interaction between them. For the RN-dS spacetime, we mainly investigate the space between black hole and cosmological horizons, i.e., the thermodynamic volume of our considered system is
\begin{eqnarray}
\widetilde{V}\equiv Q^3V=\widetilde{V}_c-\widetilde{V}_+=\frac{Q^34\pi(1-x^3)}{3x^3\phi_+^3}.\label{V}
\end{eqnarray}
The boundary of our considered system are two horizons which have different radiation temperatures. That is because the system is in gravity field, and the horizon temperatures are deformed by the warped space. Thus we regard this system as an ordinary thermodynamic system in the thermodynamic equilibrium, which is of the thermodynamic quantities ($\widetilde{T}_{eff},~\widetilde{P}_{eff},~\widetilde{V},~\widetilde{S}$ or $T_{eff},~P_{eff},~V,~S$). Next we will investigate the thermodynamical quantities of RN-dS system: such as $T_{eff},~P_{eff},~V,~S$, which are called the induced thermodynamical quantities and they are both the functions of $\phi_+$ and $x$. Here we point out the entropy is not only the sum of two horizons, it should contain the connection between two horizons which arise from the gravity effect. We assume the total entropy of this system as the following form
\begin{eqnarray}
\widetilde{S}\equiv Q^2S=\frac{Q^2\pi(1+x^2+f_0(x))}{x^2\phi_+^2},\label{S}
\end{eqnarray}
where $f_0(x)$ stands for the interplay between two horizons and it arises from the gravity effect. In the following, we mainly focus on the process of obtaining the function $f_0(x)$.

\begin{figure}[htp]
{\includegraphics[width=0.4\textwidth]{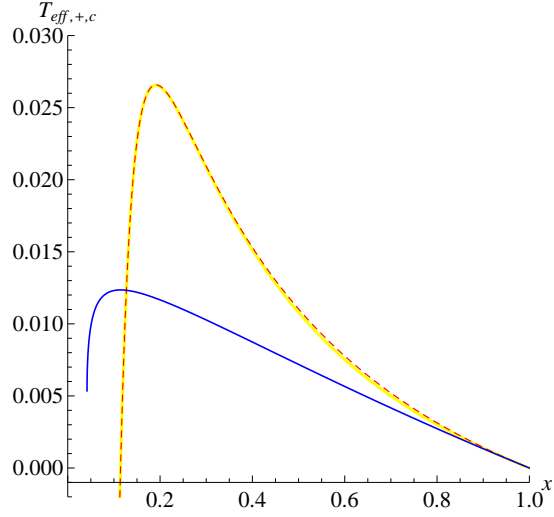}}
\caption{The pictures of $T_{eff,+,c}$ with the parameter $Q^2/l^2=\frac{1}{100}$. The yellow thick line, red thin dashed line, and blue line stand for $T_{eff}$, $T_{+}$, and $T_{c}$, respectively.}\label{Tceff}
\end{figure}

When regarding the RN-dS spacetime as an ordinary thermodynamic system in the thermodynamic equilibrium, the first law should be held on
\begin{equation}
d \widetilde{M}=\widetilde{T}_{e f f} d \widetilde{S}-\widetilde{P}_{e f f} d \widetilde{V}+\Phi_{e f f} d Q.
\end{equation}
Thus the effective temperature and pressure can be obtained from the following expressions
\begin{eqnarray}
\widetilde{T}_{e f f}&=&\frac{\partial\widetilde{M}}{\partial\widetilde{S}}\bigg|_{\widetilde{V},Q}
=\frac{\frac{\partial\widetilde{M}}{\partial\phi_+}\frac{\partial\widetilde{V}}{\partial x}-\frac{\partial\widetilde{M}}{\partial x}\frac{\partial\widetilde{V}}{\partial \phi_+}}{\frac{\partial\widetilde{S}}{\partial\phi_+}\frac{\partial\widetilde{V}}{\partial x}-\frac{\partial\widetilde{S}}{\partial x}\frac{\partial\widetilde{V}}{\partial \phi_+}}\bigg|_{Q},\label{Teff}\\
\widetilde{P}_{e f f}&=&-\frac{\partial\widetilde{M}}{\partial\widetilde{V}}\bigg|_{\widetilde{S},Q}
=\frac{\frac{\partial\widetilde{M}}{\partial\phi_+}\frac{\partial\widetilde{S}}{\partial x}-\frac{\partial\widetilde{M}}{\partial x}\frac{\partial\widetilde{S}}{\partial \phi_+}}{\frac{\partial\widetilde{V}}{\partial\phi_+}\frac{\partial\widetilde{S}}{\partial x}-\frac{\partial\widetilde{V}}{\partial x}\frac{\partial\widetilde{S}}{\partial \phi_+}}\bigg|_{Q}.\label{Peff}
\end{eqnarray}
When $\frac{Q^2}{l^2}$ satisfies eq. (\ref{ql2}), the temperature corresponding to two horizons is equal to each other. In this case, we believe that the effective temperature should be the radiation one
\begin{eqnarray}
\widetilde{T}_{e f f}=\frac{x(1+x^4)\widetilde{T}_+}{(1-x^3)\left[x(1+x)+x^2f_0+\frac{1}{2}(1-x^3)f_0'\right]}=\widetilde{T}_+.
\end{eqnarray}
As $x=0$, the interplay between two horizons should be vanished, i.e., $f_0(0)=0$. By solving above equation the interaction function of entropy reads
\begin{eqnarray}
f_0(x)=\frac{8}{5}(1-x^3)^{\frac{2}{3}}-\frac{2(4-5x^3-x^5)}{5(1-x^3)}.\label{f0}
\end{eqnarray}
Substituting eq. (\ref{f0}) into eqs. (\ref{Teff}) and (\ref{Peff}), the effective temperature and effective pressure are satisfied the following forms as shown in Ref. \cite{Zhang2020}
\begin{eqnarray}
0&=&Q\widetilde{T}_{eff}f_1(x)-f_2(x)\phi_++f_3(x)\phi_+^3,~~~~\text{or}~~~~0=T_{eff}f_1(x)-f_2(x)\phi_++f_3(x)\phi_+^3\label{T},\\
0&=&Q^2\widetilde{P}_{eff}f_4(x)+f_5(x)\phi_+^2+f_6(x)\phi_+^4, ~~~\text{or}~~~~0=P_{eff}f_4(x)+f_5(x)\phi_+^2+f_6(x)\phi_+^4\label{P}
\end{eqnarray}
with
\begin{eqnarray}
f_1(x)&=&\frac{4\pi(1+x^4)}{1-x},~~f_2(x)=\left[(1-3x^2)(1+x+x^2)+4x^3(1+x)\right],f_3(x)=(1+x+x^2)(1+x^4)-2x^3,~~\nonumber\\
f_4(x)&=&\frac{8\pi(1+x^4)}{x(1-x)},~~~
f_5(x)=kx(1+x)(x+f_0'/2)-\frac{k(1+2x)(1+x^2+f_0)}{1+x+x^2},~~~\nonumber\\
f_6(x)&=&\frac{(1+2x+3x^2)(1+x^2+f_0)}{1+x+x^2}-x(1+x)(1+x^2)(x+f_0'/2).\nonumber
\end{eqnarray}
In order to understand the temperatures on two horizons and the effective temperature with different $Q^2/l^2$, we present the picture of them in Fig. \ref{Tceff}.
Adopting $Q,~\widetilde{S},~\widetilde{V}$ as the independent parameters of $\widetilde{M}$ and considering the units of these quantities, the smarr formula reads
\begin{eqnarray}
\widetilde{M}=2\widetilde{T}_{eff}\widetilde{S}-3\widetilde{P}_{eff}\widetilde{V}+Q\Phi_{eff}.
\end{eqnarray}
In the same way for the reduced quantities $T_{eff},~P_{eff},~S,~V,~M$, the reduced smarr formula becomes
\begin{eqnarray}
M=2T_{eff}S-3P_{eff}V.
\end{eqnarray}
These results can be verified by substituting eqs. (\ref{T}), (\ref{S}), (\ref{V}), (\ref{P}), and (\ref{Mreduced}) into the above equation.

\begin{figure}[htp]
\subfigure[$C_{Q^2/l^2}-x$]{\includegraphics[width=0.4\textwidth]{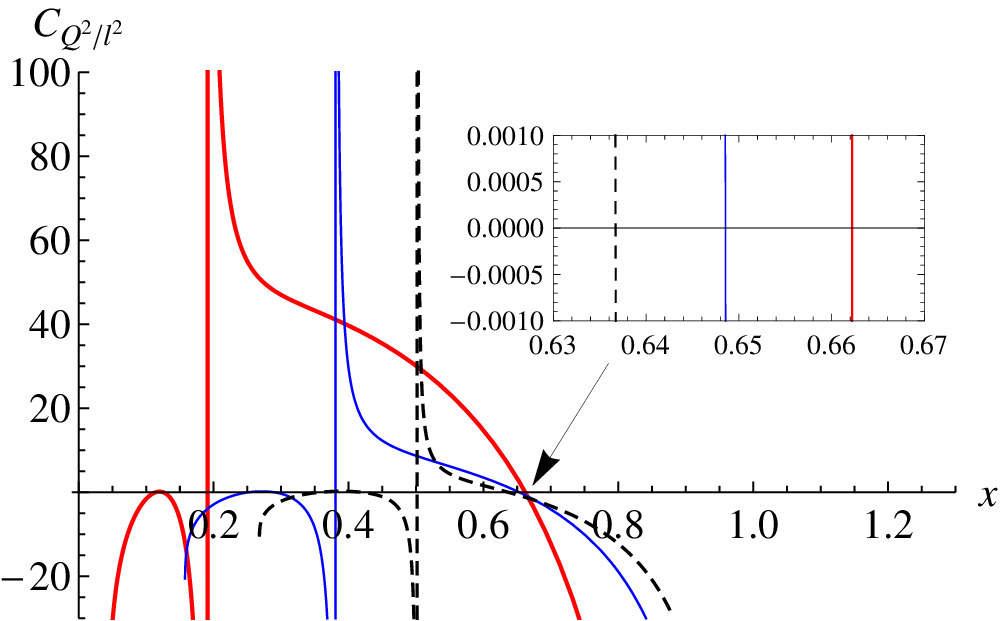}\label{Cql2}}~~~~
\subfigure[$C_{P_{eff}}-x$]{\includegraphics[width=0.4\textwidth]{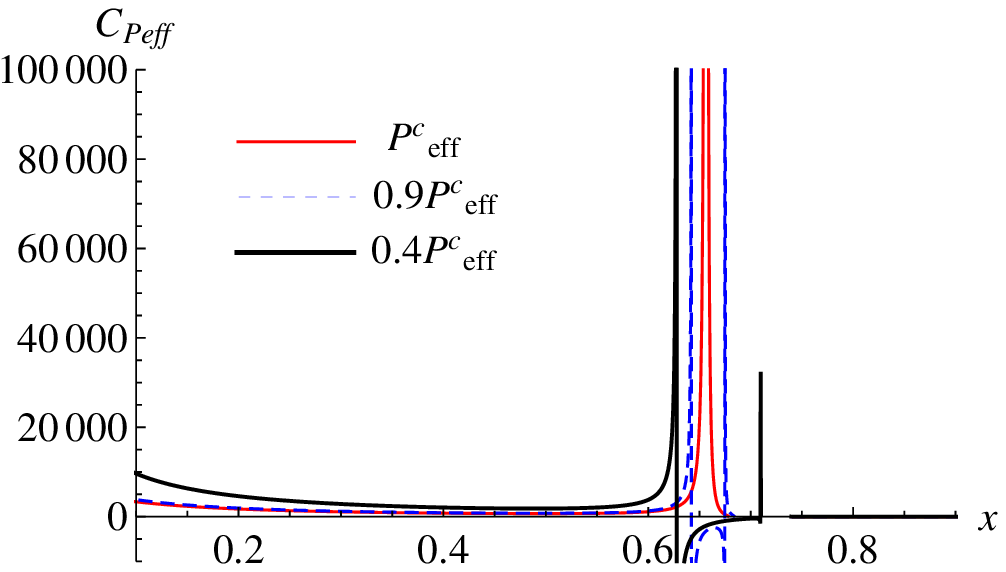}\label{Cp0}}
\caption{The pictures of $C_{Q^2/l^2}$ and $C_{P_{eff}}$. The parameter of $Q^2/l^2$ is set to $\frac{1}{100}$ (red thick line), $\frac{1}{30}$ (blue thin line), and $\frac{1}{20}$ (black thin dashed line).}\label{C}
\end{figure}

\begin{figure}[htp]
\includegraphics[width=0.4\textwidth]{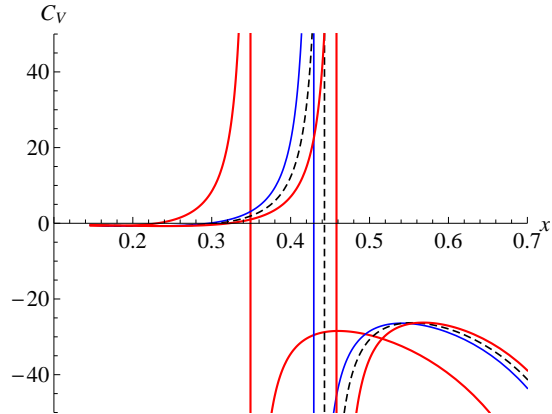}
\caption{The pictures of $C_{V}$ with different volumes. The parameter of $V$ is set to $V_c$ (red thick line), $0.9V_c$ (blue thin line), and $0.5V_c$ (black thin dashed line).}\label{C0}
\end{figure}

\begin{figure}[htp]
\subfigure[$\alpha-x$]{\includegraphics[width=0.4\textwidth]{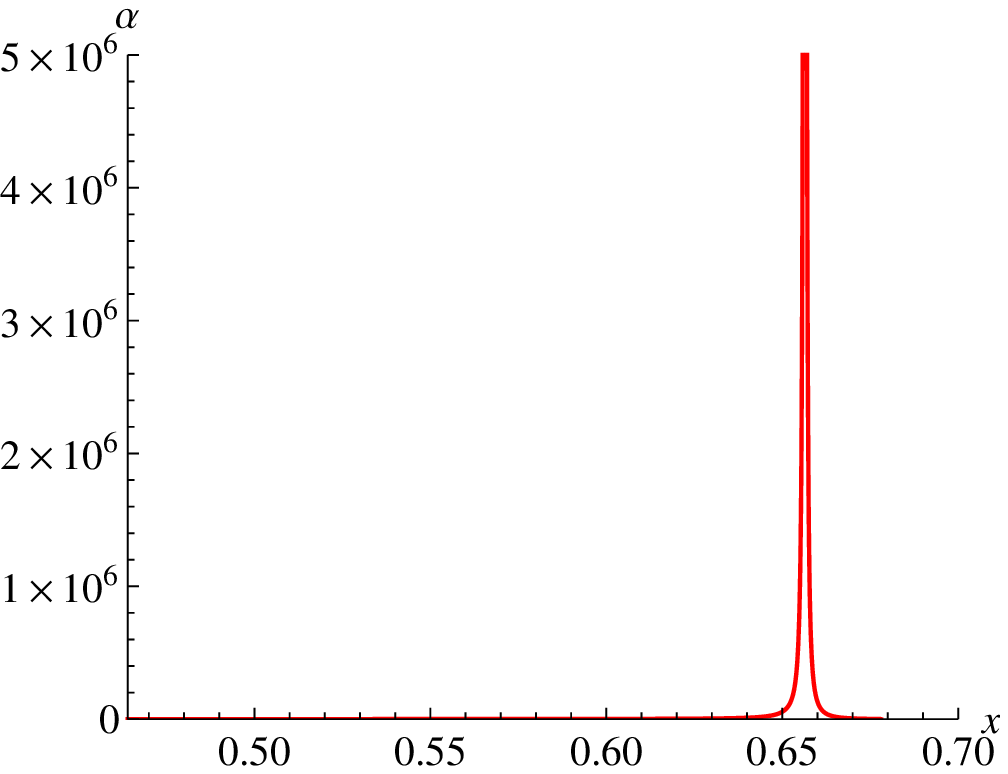}\label{alpha-x}}~~~~
%\subfigure[$\beta-x$]{\includegraphics[width=0.4\textwidth]{beta-x.eps}\label{beta-x}}\\
\subfigure[$\kappa-x$]{\includegraphics[width=0.4\textwidth]{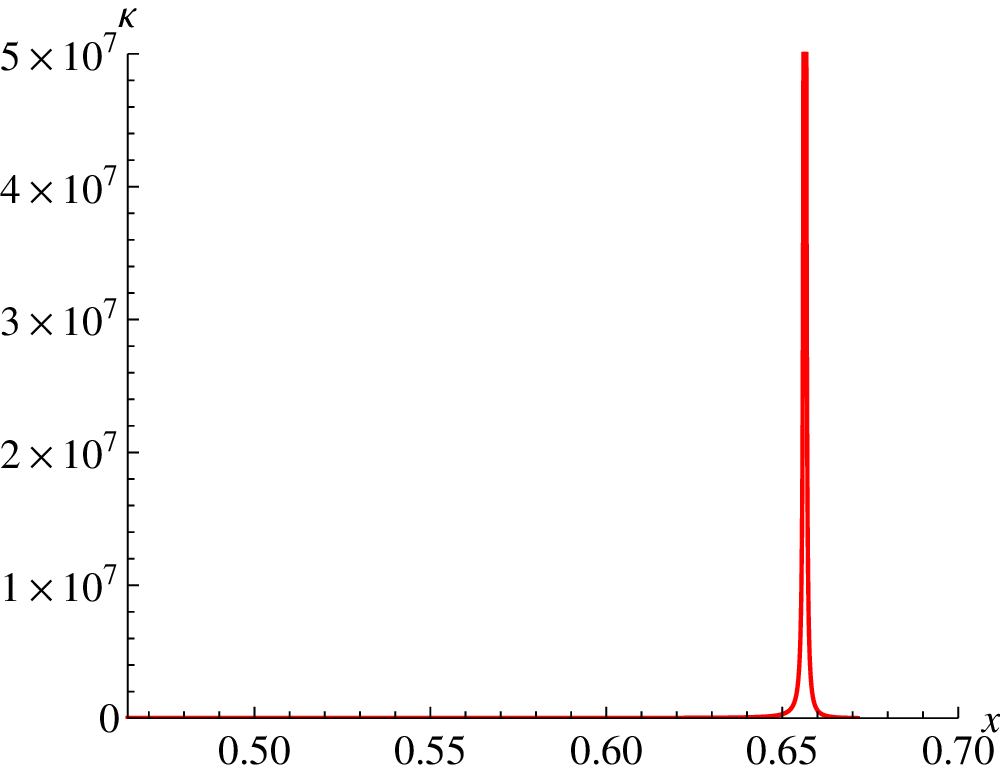}\label{kappa-x}}
\caption{The pictures of three coefficients with two horizons' radio $x$ nearby the critical point.}\label{coefficientsx}
\end{figure}

The critical point denoting a second-order phase transition is determined by the following equations
\begin{eqnarray}
\frac{\partial\widetilde{P}}{\partial\widetilde{V}}\bigg|_{\widetilde{T}_{eff},Q}=\frac{\partial^2\widetilde{P}}{\partial\widetilde{V}^2}\bigg|_{\widetilde{T}_{eff},Q}=0,~~~~
\text{or} ~~~~\frac{\partial P}{\partial V}\bigg|_{T_{eff}}=\frac{\partial^2P}{\partial V^2}\bigg|_{T_{eff}}=0.
\end{eqnarray}
With eqs. (\ref{V}), (\ref{f0}), (\ref{T}), and (\ref{P}), the critical reduced thermodynamical quantities are
\begin{eqnarray}
x^c=0.65646,~~\phi^c=0.378244,~~T_{eff}^c=0.00863395,~~P_{eff}^c=0.000583686,~~V_c=196.214,~~S_c=68.2089.
\end{eqnarray}
Performing isobaric, isothermal, and isovolume processes for the system and solving eqs. (\ref{T}), (\ref{P}), and (\ref{V}), the physical potentials on RN-dS black hole horizon $\phi_+$ become
\begin{eqnarray}
\phi_+&=&\phi_p=\sqrt{\frac{2P_{eff}f_4}{-f_5+\sqrt{f_5^2-4P_{eff}f_4f_6}}},\label{phip}\\
\phi_+&=&\phi_t=\sqrt{\frac{4f_2}{3f_3}}\cos\left(\theta+\frac{4\pi}{3}\right),~
\theta=\frac{1}{3}\arccos\left(\frac{-T_{eff}f_1\sqrt{27f_2f_3}}{2f_2^2}\right),\label{phit}\\
\phi_+&=&\phi_v=\left(\frac{4\pi[1-x^3]}{3x^3V}\right)^{1/3}.
\end{eqnarray}
For a real RN black hole embedded in a dS spacetime, since the black hole horizon is not coincide with the cosmological horizon, the value of $x$ cannot be one. In order to solving this problem, we study the heat capacity at the constant $Q^2/l^2$ and at the constant effective pressure $P_{eff}$, respectively, i.e., for a real stable RN-dS spacetime the heat capacities should be positive. The heat capacities at the constant $Q^2/l^2$ and $P_{eff}$ are
\begin{eqnarray}
\widetilde{C}_{Q^2/l^2}
=\widetilde{T}_{eff}\left(\frac{\partial\widetilde{S}}{\partial\widetilde{T}_{eff}}\right)\bigg|_{Q^2/l^2}
\equiv Q^2C_{Q^2/l^2},~~~~
\widetilde{C}_{P_{eff}}
=\widetilde{T}_{eff}\left(\frac{\partial\widetilde{S}}{\partial\widetilde{T}_{eff}}\right)\bigg|_{P_{eff}}
\equiv Q^2C_{P_{eff}}.\nonumber
\end{eqnarray}
Here the reduced heat capacities read
\begin{eqnarray}
C_{Q^2/l^2}&=&T_{eff}\left(\frac{\partial S}{\partial T_{eff}}\right)\bigg|_{Q^2/l^2},\label{Cql}\\
C_{P_{eff}}&=&T_{eff}\left(\frac{\partial S}{\partial T_{eff}}\right)\bigg|_{P_{eff}}.\label{Cp}
\end{eqnarray}
Combining eqs. (\ref{ql2}), (\ref{S}), (\ref{f0}), (\ref{T}) and (\ref{phip}), the pictures of heat capacities with different values of $Q^2/l^2$ and $P_{eff}$ are shown in Fig. \ref{C}.
As what we have mentioned, from Fig. \ref{Cql2} we can see that the value of $x$ for a stable RN-dS spacetime is from $x_{min}$ to $x_{max}$, where the maximum and minimum of $x$ is related with $Q^2/l^2$. Note that for the given $Q^2/l^2$ the minimum of $x$ in eq. (\ref{Cql}) is bigger than that in eq. (\ref{x}). Therefore, the thermodynamical property of RN-dS spacetime will be investigated in a certain range of $x$ obtained from the view point of $C_{Q^2/l^2}$. In Fig. \ref{Cp0}, $C_{P_{eff}}$ is divergent at the critical point and at two different values of $x$ when $P_{eff}<P_{eff}^c$. In the next, we will present the two divergent location of $C_{P_{eff}}$ when $P_{eff}<P_{eff}^c$ are just the ratios between two horizons' potentials of two coexistent phases. Similarly, the reduced heat capacity with the constant volume becomes
\begin{eqnarray}
C_{V}&=&T_{eff}\left(\frac{\partial S}{\partial T_{eff}}\right)\bigg|_{V}.\label{Cv}
\end{eqnarray}
From eqs. (\ref{V}), (\ref{S}), (\ref{f0}), and (\ref{T}), the plots of the heat capacity with different volumes are exhibited in Fig. \ref{C0}.
In addition there are some coefficients to probe the critical behavior of a thermodynamical system, i.e., the isobaric expansion coefficient and isothermal compression coefficient
\begin{eqnarray}
\widetilde{\alpha}=\frac{1}{\widetilde{V}}\left(\frac{\partial\widetilde{V}}{\partial\widetilde{T}_{eff}}\right)\bigg|_{\widetilde{P}_{eff}},~~
%\widetilde{\beta}=\frac{1}{\widetilde{P}_{eff}}\left(\frac{\partial\widetilde{P}_{eff}}{\partial\widetilde{T}_{eff}}\right)\bigg|_{\widetilde{V}},~~
\widetilde{\kappa}=-\frac{1}{\widetilde{V}}\left(\frac{\partial\widetilde{V}}{\partial\widetilde{P}_{eff}}\right)\bigg|_{\widetilde{T}_{eff}}.\nonumber
\end{eqnarray}
The introduced isobaric expansion coefficient and isothermal compression coefficient have the following forms
\begin{eqnarray}
\widetilde{\alpha}&\equiv& Q\alpha,~~~~\alpha=\frac{1}{V}\left(\frac{\partial V}{\partial T_{eff}}\right)\bigg|_{P_{eff}},\\
%\widetilde{\beta}&\equiv& Q\beta,~~~~\beta=\frac{1}{P_{eff}}\left(\frac{\partial P_{eff}}{\partial T_{eff}}\right)\bigg|_{V},\\
\widetilde{\kappa}&\equiv&
\frac{\kappa}{Q^2},~~~~\kappa=-\frac{1}{V}\left(\frac{\partial V}{\partial P_{eff}}\right)\bigg|_{T_{eff}}.
\end{eqnarray}
Their behaviors nearby the critical point are exhibited in Fig. \ref{coefficientsx}.

\section{First-order phase transition}
\label{three}
\begin{figure}[htp]
\subfigure[$G_{P_{eff}}-T_{eff}$]{\includegraphics[width=0.4\textwidth]{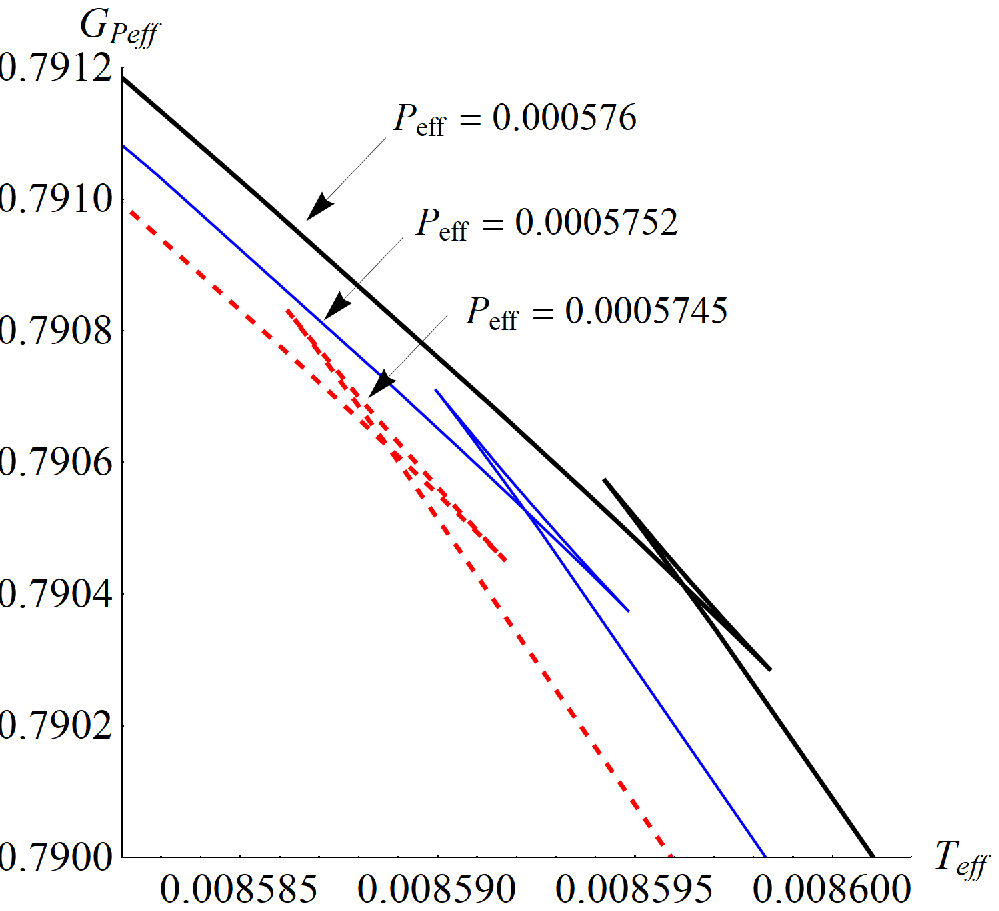}\label{CT}}~~~~
\subfigure[$G_{T_{eff}}-P_{eff}$]{\includegraphics[width=0.4\textwidth]{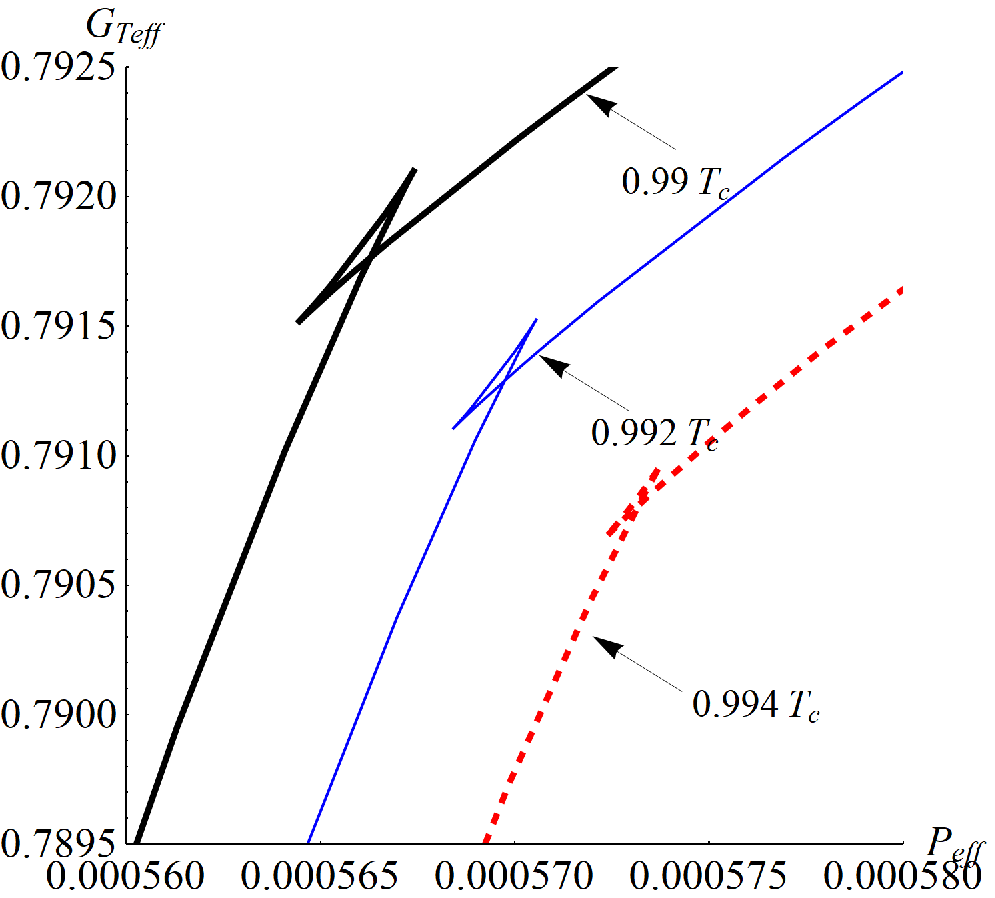}\label{CP}}~~~~\\
\subfigure[$G-T_{eff}-P_{eff}$]{\includegraphics[width=0.4\textwidth]{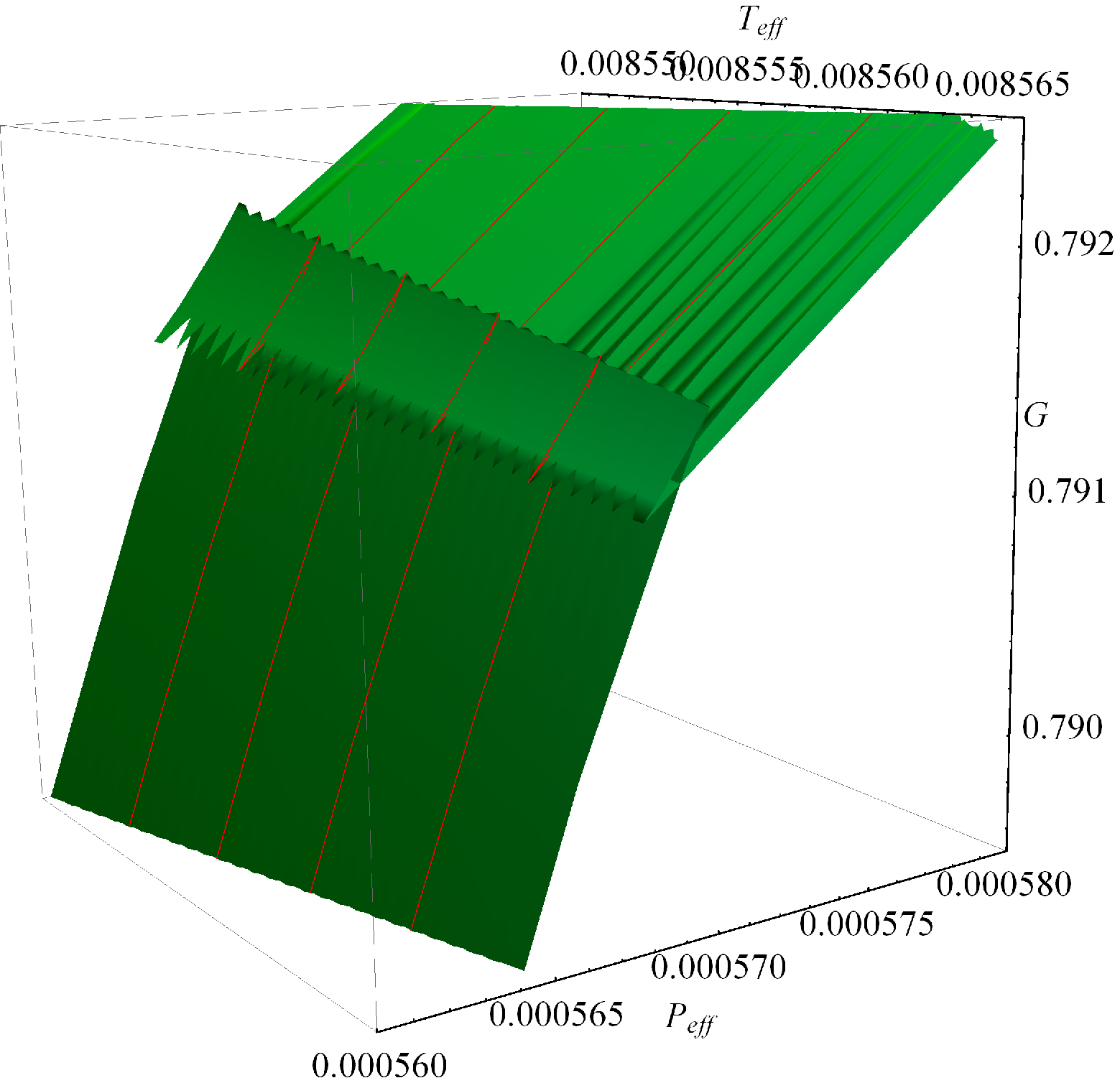}\label{CTP}}~~~~
\caption{The swallow tail behaviors of Gibbs free energy nearby the critical point in $2$-dimension and $3$-dimension .}\label{G}
\end{figure}

\begin{figure}[htp]
\subfigure[$T_{eff}=0.995T_{eff}^c$]{\includegraphics[width=0.4\textwidth]{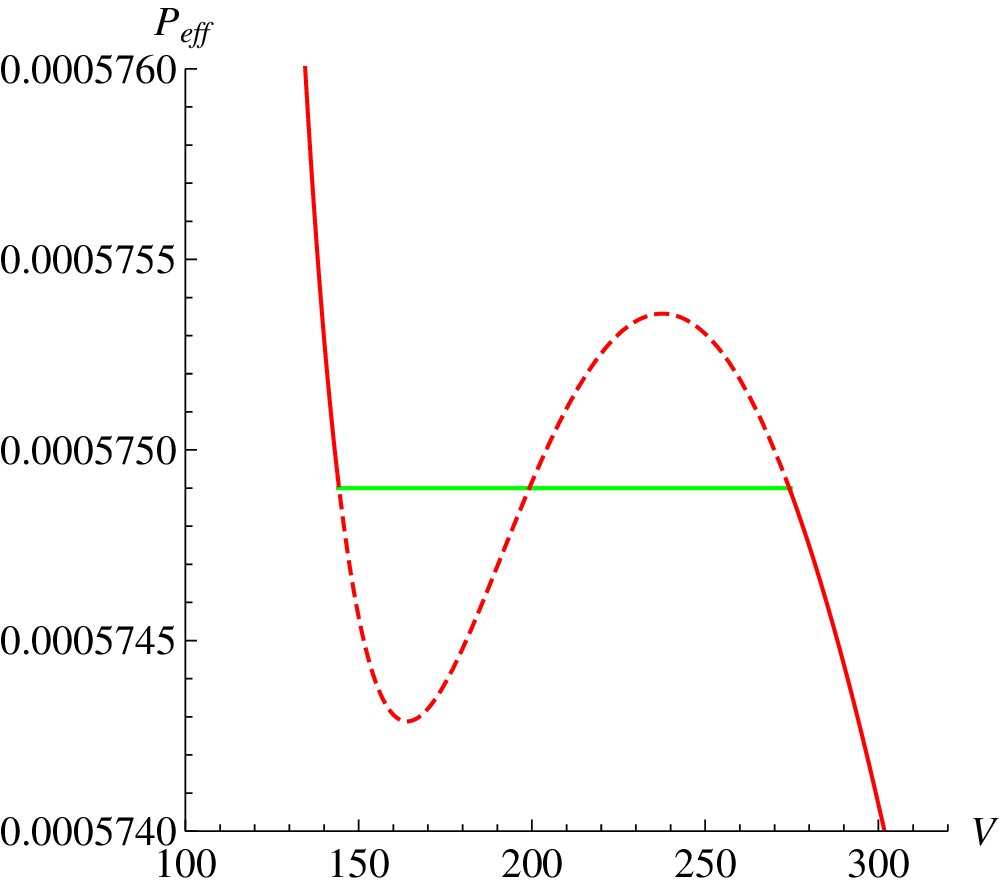}\label{PV}}~~~~
\subfigure[$P_{eff}=0.985P_{eff}^c$]{\includegraphics[width=0.4\textwidth]{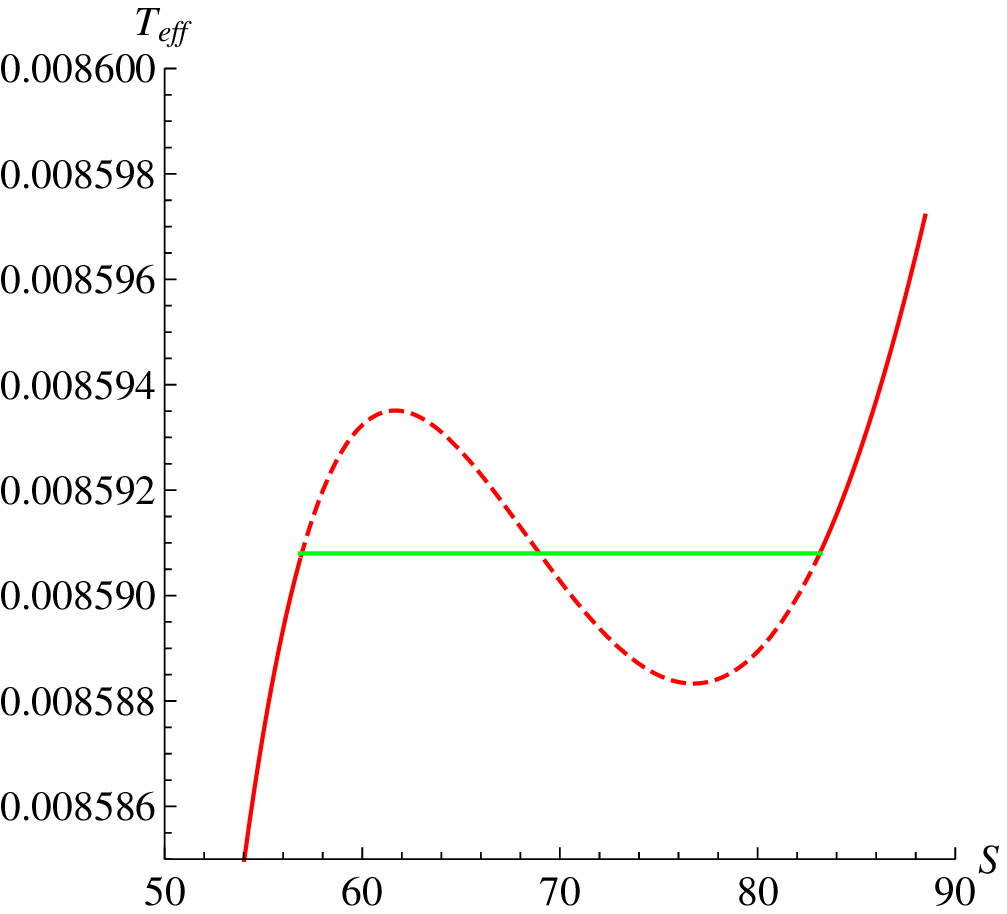}\label{TS}}~~~~
\caption{The equal-area law in the phase diagrams of $P_{eff}-V$ and $T_{eff}-S$ .}\label{EAL}
\end{figure}

In this section, we would like to study the first-order phase transition in the canonical ensemble. From the effective thermodynamical quantities of RN-dS spacetime, the thermodynamical state equation can be written as
\begin{eqnarray}
F(\widetilde{T}_{eff},~\widetilde{P}_{eff},\widetilde{S},~\widetilde{V},~Q)=0,~~~~\text{or}~~~~
F(T_{eff},~P_{eff},S,~V,)=0.
\end{eqnarray}
Gibbs free energy of this system reads
\begin{eqnarray}
\widetilde{G}\equiv QG=\widetilde{M}-\widetilde{T}_{eff}\widetilde{S}+\widetilde{P}_{eff}\widetilde{V}=Q(M-T_{eff}S+P_{eff}V)
\end{eqnarray}
with the reduced mass parameter
\begin{eqnarray}
M=\frac{1}{2\phi_+}\left(\frac{1-x^2}{1-x^3}+\frac{1-x^4}{1-x^3}\phi_+^2\right). \label{Mreduced}
\end{eqnarray}
Combining eqs. (\ref{V}), (\ref{S}), (\ref{f0}), (\ref{T}), (\ref{P}), (\ref{phip}), and (\ref{phit}), the behaviors of Gibbs free energy nearby the critical point in $2$-dimension and $3$-dimension are shown in Fig. \ref{G}.
In addition, the first-order phase transition point can also be obtained by the construction of equal-area law in different phase diagrams. Here we exhibit the first-order phase diagrams of $P_{eff}-V$ and $T_{eff}-S$ in Fig. \ref{EAL}.

\begin{figure}[htp]
\includegraphics[width=0.4\textwidth]{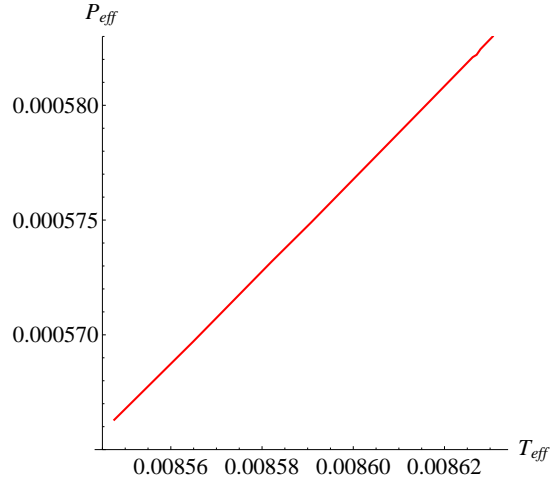}
\caption{The coexistent curve in the phase diagrams of $T_{eff}-P_{eff}$ nearby the critical point.}\label{TP}
\end{figure}

In fact, the first-order phase transition point is a coexistent state of two different phases in RN-dS spacetime. All these points form one coexistent curve. Although there is no the analytic form for the curve, it can be described the following behavior with the numerical results in Fig. \ref{TP}. And the thermodynamic quantities of two phases with different first-order phase transition points are presented in Tab. \ref{tcp}.

\begin{table}[htbp]
\centering
\caption{the thermodynamic quantities of two coexistent phases for $T_{eff}^0\leq T_{eff}^c$}
\begin{tabular}{|c|c|c|c|c|c|c|c|}
\hline
\centering
$T_{eff}^0$~~&$P_{eff}^0$~~& $x_{1}$~~&$x_{2}$~~&$\phi_{+1}$~~&$\phi_{+2}$~~&$R_1$~~&$R_2$ \\ \hline
~~$0.99T_{eff}^c$ ~~&0.0005663~~&0.6399~~&0.6707~~&0.3343~~&0.4235~~&0.759027~~&0.808721\\ \hline
$0.992T_{eff}^c$~~&0.0005697~~&0.6416~~&0.6693~~&0.3386~~&0.4186~~&0.84761~&0.903014\\ \hline
$0.994T_{eff}^c$~~&0.0005732~~& 0.6439~~& 0.6677~~& 0.3442~~& 0.4133~~& 0.994135~~& 1.01518\\ \hline
$0.995T_{eff}^c$~~&0.0005749~~& 0.6448~~&0.6668~~& 0.3466~~& 0.4102~~& 1.05345~~& 1.09469\\ \hline
$0.996T_{eff}^c$~~&0.000576653~~& 0.6462~~&0.6657~~&0.3502~~&0.4064~~& 1.16212~~& 1.20905\\ \hline
$0.997T_{eff}^c$~~&0.000578405~~& 0.6467~~&0.6646~~& 0.3538~~& 0.4029~~& 1.29083~~& 1.3191\\ \hline
$0.998T_{eff}^c$~~&0.0005801595~~& 0.6493~~& 0.6631~~& 0.3584~~& 0.3982~~& 1.46474~~& 1.49429\\ \hline
$0.999T_{eff}^c$~~&0.0005819201~~& 0.6514~~& 0.6612~~& 0.3639~~& 0.3922~~& 1.80019~~& 1.82641\\ \hline
$0.9993T_{eff}^c$~~&0.0005824486~~& 0.6522~~& 0.6604~~& 0.3662~~& 0.3899~~& 1.92668~~& 1.94651\\ \hline
$0.9995T_{eff}^c$~~&0.000582804~~& 0.6529~~& 0.6598~~& 0.3681~~& 0.3881~~& 2.10521~~& 2.08337\\ \hline
$T_{eff}^c$~~&$P_{eff}^c=0.000583686$~~&$x^c=0.65646$~~&$x^c=0.65646$~~&$\phi_+^c=0.378244$
~~&$\phi_+^c=0.378244$~~&$R^c=3.12592$ ~~&$R^c=3.12592$ \\ \hline
\end{tabular}\label{tcp}
\end{table}

Now we exploit Ehrenfest's scheme in order to understand the nature of the phase transition, which basically consists of a pair of equations known as Ehrenfest's equations and second kind. For a standard thermodynamic system these equations may be written as \cite{Banerjee2011,Banerjee2011a,Abbasvandi2019}
\begin{eqnarray}
\frac{\partial P_{eff}}{\partial T_{eff}}\bigg|_S&=&\frac{C_{P_{eff2}}-C_{P_{eff1}}}{T_{eff}V(\alpha_2-\alpha_1)}=\frac{\Delta C_{P_{eff}}}{T_{eff}V\Delta\alpha},\\
\frac{\partial P_{eff}}{\partial T_{eff}}\bigg|_V&=&\frac{\alpha_2-\alpha_1}{\kappa_{T_{eff2}}-\kappa_{T_{eff1}}}=\frac{\Delta \alpha}{\Delta\kappa_{T_{eff}}}.
\end{eqnarray}
For a genuine second-order phase transition both of these equations have to be satisfied simultaneously. With the expressions $\frac{\partial P_{eff}}{\partial T_{eff}}\big|_S=\frac{\partial S}{\partial V}\big|_{P_{eff}}$ and $\frac{\partial P_{eff}}{\partial T_{eff}}\big|_V=\frac{\partial S}{\partial V}\big|_{T_{eff}}$, the Prigogine-Defay (PD) ratio \cite{Jackle1986} of the RN-dS spacetime may be found to be
\begin{eqnarray}
\Pi=\frac{\partial S}{\partial V}\bigg|_{P_{eff}}\bigg/\frac{\partial S}{\partial V}\bigg|_{T_{eff}}=1.
\end{eqnarray}
Hence the phase transition occurring at $T_{eff}^c$ is a second-order equilibrium transition, which is similar to that of AdS black hole. This is true in spite of the fact that the phase transition curves are smeared and divergent near the critical point.

\section{Joule-Thomson expansion}
\label{four}
\begin{figure}[htp]
\includegraphics[width=0.4\textwidth]{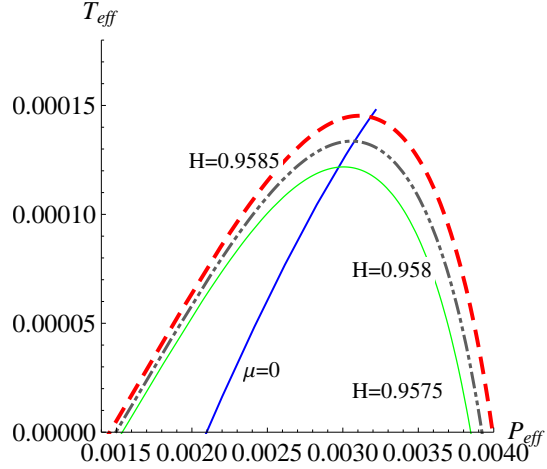}
\caption{The isenthalphic and inverse curves of the RN-dS spacetime.}\label{PTH}
\end{figure}

More recently, the authors of \cite{Okcu2017} have investigated the Joule-Thomson (JT) expansion for AdS charged black holes with the aim to confront the resulting features with those of Van der Waals fluids. The extension to rotating-AdS black hole \cite{Okcu2018} and the charged black hole solution in the presence of the quintessence field \cite{Ghaffarnejad2018} have also been considered. JT expansion \cite{Perry1934} is a convenient isoenthalpic tool that a thermal system exhibits with a thermal expansion. It is worth noting that when expanding a thermal system with a temperature $T$, the pressure always decreases yielding a negative sign to $\partial P$. In this section, we will investigate the Joule-Thomson expansion of the RN-dS spacetime. In Joule-Thomson expansion for the Van der Waals system as well as AdS black holes, gas/black hole phase is passed at high pressure through a porous plug or small value in the low-pressure section of an adiabatic tube, and the enthalpy remains constant during the expansion. The expansion is characterized by a change in temperature relative to pressure. In addition the enthalphy can be used to define the non-equilibrium states. Thus we can also introduce it to describe the RN-dS spacetime. The Joule-Thomson coefficient, which can describe the expansion process, is read as
\begin{eqnarray}
\widetilde{\mu}=\frac{\partial\widetilde{T}_{eff}}{\partial\widetilde{P}_{eff}}\bigg|_{\widetilde{H}},
\end{eqnarray}
where the enthalphy is
\begin{eqnarray}
\widetilde{H}=\widetilde{M}+\widetilde{P}_{eff}\widetilde{V}.
\end{eqnarray}
For similarity, we introduce the reduced enthalphy and the Joule-Thomson coefficient as
\begin{eqnarray}
\widetilde{H}&\equiv&QH,~~~~H=M+PV, \label{H}\\
\widetilde{\mu}&\equiv&Q\mu,~~~~~~\mu=\frac{\partial T_{eff}}{\partial P_{eff}}\bigg|_{H}=\frac{V}{C_{P_{eff}}}\left(T_{eff}\alpha-1\right).
\end{eqnarray}
For an isenthalphy for this system, the relation between potential and two horizons' ratio $x$ reads
\begin{eqnarray}
\phi_+=\phi_{+h}=\frac{H}{F_1}\left(1\pm\sqrt{1-\frac{F_1F_2}{H^2}}\right)\label{phih}
\end{eqnarray}
with
\begin{eqnarray}
F_1=\frac{1-x^4}{1-x^3}-\frac{8\pi(1-x^3)f_6}{3x^3f_4},~~~F_2=\frac{1-x^2}{1-x^3}-\frac{8\pi(1-x^3)f_5}{3x^3f_4}.
\end{eqnarray}
When $\alpha T_{eff}=1$, i.e., $\mu=0$, we can obtain
\begin{eqnarray}
\phi_{+0}=\sqrt{\frac{-N_2\left(1\mp\sqrt{1-4N_1N_3/N_2^2}\right)}{2N_1}}\label{phi0}
\end{eqnarray}
with
\begin{eqnarray}
N_1&=&\frac{4f_6}{f_4}\left[\frac{f_3}{f_1}\right]'-\frac{6f_3}{f_1}\left[\frac{f_6}{f_4}\right]'-\frac{4}{x(1-x^3)}\frac{f_3f_6}{f_1f_4},~~~\nonumber\\
N_2&=&\frac{4f_2}{f_1}\left[\frac{f_6}{f_4}\right]'-\frac{6f_3}{f_1}\left[\frac{f_5}{f_4}\right]'+\frac{2}{x(1-x^3)}\left(\frac{2f_2f_6}{f_1f_4}-\frac{f_3f_5}{f_1f_4}\right)
-\frac{4f_6}{f_4}\left[\frac{f_2}{f_1}\right]'-\frac{2f_5}{f_4}\left[\frac{f_3}{f_1}\right]',~~~~\nonumber\\
N_3&=&\frac{4f_2}{f_1}\left[\frac{f_5}{f_4}\right]'-\frac{2f_5}{f_4}\left[\frac{f_2}{f_1}\right]'-\frac{2}{x(1-x^3)}\frac{f_2f_5}{f_1f_4}.~~~~\nonumber
\end{eqnarray}
To better investigate the Joule-Thomson expansion, by substituting eqs. (\ref{phih}) and (\ref{phi0}) into eqs. (\ref{T}) and (\ref{P}) respectively, the isoenthalpic and inversion curves of this system are depicted in Fig. \ref{PTH}. The result shows that the inversion curve divide the isoenthalpic one into two parts: one is the cooling phenomena with the positive slope of the $P-T$ curves, the other is the heating process with the negative slope of the $P-T$ curves.

\section{Landau' theory of continuous phase transition}
\label{five}
For an ordinary thermodynamic system, the second-order derivative of Gibbs free energy, heat capacity, isobaric expansion and isothermal compression coefficients have a peak or emerge a jump nearby the critical point. Additionally, the power functions also are used to describe the critical behavior, whose powers are called the critical exponents. Based on the experiments of various ferromagnetic systems and the adoption of the spontaneous magnetization as the order parameter, there exist some results nearby critical point with $\bar{t}=T-T_c$:
\begin{itemize}
  \item{Spontaneous magnetization satisfies $M'\propto(-\bar{t})^{\beta'},~\bar{t}\rightarrow0^{-}$, and it equals to zero when above the critical temperature;}
  \item{Magnetic susceptibility obeys $\chi\propto \bar{t}^{-\gamma'},~\bar{t}\rightarrow0^+;~\chi\propto \bar{t}^{-\gamma''},~\bar{t}\rightarrow0^-$;}
  \item{Magnetization and applied magnetic field satisfies $M'\propto\widetilde{H}^{1/\delta},~\bar{t}\rightarrow0$;}
  \item{Heat capacity without applied magnetic field of ferromagnetic matters obey the following expressions\\ ~~~~~~~~~~~~~~~~~~~~~$C\propto \bar{t}^{-\alpha'},~\bar{t}>0;~C\propto (-\bar{t})^{-\alpha''}~\bar{t}>0.$}
  \item{Critical exponents read $\beta'=1/2,~\gamma'=\gamma''=1,~\delta=3,~\alpha'=\alpha''=0.$}
\end{itemize}
Subsequently, people found that for various fluid systems the critical behavior is the same as well as ferromagnetic systems. Furthermore, the critical behavior of systems with completely different physical characteristics have extremely similar characteristics compared with fluid and ferromagnetic systems. That indicates the critical phenomena has a certain universality. Based on the investigation of AdS black holes, the critical powers calculated theoretically are consistent with that of ferromagnetic systems. For the RN-dS spacetime, the critical phenomena should be also focused. However, it is different from AdS black holes and ferromagnetic systems, and its phase transition is determined by the ratio of electric potentials between two horizons as shown in Sec. \ref{four}. Hence, the explore of critical phenomena for RN-dS spacetime will help us better understand its thermodynamic properties.

For the Landau' theory of continuous phase transition, the phase transition is accompanied by a change in the degree of order and symmetry of system. By regarding the RN-dS spacetime as a thermodynamic system which is composed of internal molecules, whether there also exist the changes of internal symmetry and order when phase transition emerges becomes a question of general concern. Though the investigation shown in Sec. \ref{three}, we found that when the phase transition emerges, for the system in the higher-potential black hole phase, the internal molecules have a certain orientation under the action of strong electric potential, and are in an orderly and high symmetric state; while for the system in the low-potential black hole phase they are in a low-order and low symmetric state. Increasing temperature above the critical one, the thermal motion of internal molecules increases, making their orientation tend to zero.

In order to explain the nature of phase transition of RN-dS spacetime, a new order parameter $\eta$ related to electric potential should be introduced to describe the degree of order. When $T_{eff}<T_{eff}^c$, the electric potentials on RN-dS black hole and cosmological horizons both have two different values: $\phi_{+1,+2}$ and $\phi_{c1,c2}$. With the definition $y\equiv\phi_{+1}/\phi_{+2}$, the difference of the electric potentials on black hole horizon for two black hole phases becomes $\bigtriangleup\phi_+=\phi_{+1}(1-y)$. As $T_{eff}\rightarrow T_{eff}^c$, $y\rightarrow 1$, $\eta\rightarrow 0$. Since $\bigtriangleup\phi_+=0$ with $T_{eff}\geq T_{eff}^c$ and $\bigtriangleup\phi_+\neq0$ with $T_{eff}\leq T_{eff}^c$, $\bigtriangleup\phi_+$ can be adopted as the order parameter: $\eta\equiv\bigtriangleup\phi_+$.

Because the order parameter is a small quantity nearby the critical point and the system is symmetry for $\eta\leftrightarrows-\eta$, for the simplicity the Gibbs free energy $G(T_{eff},P_{eff})$ can be extended to the fourth order in the power series of the order parameter as following form
\begin{eqnarray}
G(T_{eff},P_{eff},\eta)=G_0(T_{eff},P_{eff})+A(T_{eff},P_{eff})\eta^2+B(T_{eff},P_{eff})\eta^4.
\end{eqnarray}
The corresponding generalized force $\xi$ reads
\begin{eqnarray}
\xi=2A\eta+4B\eta^3.
\end{eqnarray}
Hence the critical exponent $\delta=3$. For the system in the full disorder state: $\xi=0$, by solving the above equation we have
\begin{eqnarray}
\eta=0,~~\eta=\pm\sqrt{-\frac{A}{2B}}.
\end{eqnarray}
By analyzing the above solutions we can find that when $B>0$ the system will be in a stable disorder state with $\eta=0,~A>0$; while it is in a stable order state with $\eta=\pm\sqrt{-\frac{A}{2B}},~A<0$. The exponent $\beta'$ equals $1/2$. Because of $A(T_{eff}^c,P_{eff}^c)=0$, we can assume
\begin{eqnarray}
A(T_{eff},P_{eff})\approx A_1(T_{eff}^c,P_{eff}^c)(T_{eff}-T_{eff}^c)+A_2(T_{eff}^c,P_{eff}^c)(P_{eff}-P_{eff}^c)
\end{eqnarray}
with the positive coefficients $A_1(T_{eff}^c,P_{eff}^c)$ and $A_2(T_{eff}^c,P_{eff}^c)$. For $T_{eff}>T_{eff}^c$ and $P_{eff}=P_{eff}^c$, the RN-dS spacetime is in a stable disorder state, otherwise it is in a stable order one when $T_{eff}<T_{eff}^c$ and $P_{eff}=P_{eff}^c$.

Nearby the critical point, the Gibbs free energy, entropy, and volume are continuous and have the following forms approximately
\begin{eqnarray}
G&=&G_0(T_{eff},P_{eff})+[A_1(T_{eff}^c,P_{eff}^c)(T_{eff}-T_{eff}^c)+A_2(T_{eff}^c,P_{eff}^c)(P_{eff}-P_{eff}^c)]\eta^2,\nonumber\\
S&=&-\frac{\partial G}{\partial T_{eff}}=S_0+A_1(T_{eff}^c,P_{eff}^c)\frac{A_1(T_{eff}^c,P_{eff}^c)(T_{eff}-T_{eff}^c)+A_2(T_{eff}^c,P_{eff}^c)(P_{eff}-P_{eff}^c)}{2B},\nonumber\\
V&=&\frac{\partial G}{\partial P_{eff}}=V_0+A_2(T_{eff}^c,P_{eff}^c)\frac{A_1(T_{eff}^c,P_{eff}^c)(T_{eff}-T_{eff}^c)+A_2(T_{eff}^c,P_{eff}^c)(P_{eff}-P_{eff}^c)}{2B}.\nonumber
\end{eqnarray}
The heat capacities and isothermal compression coefficient at the critical point are
\begin{eqnarray}
C_{P_{eff}^c}&=&C_{P_{eff}^c0}+\frac{A_1^2(T_{eff}^c,P_{eff}^c)}{2B}T_{eff}^c,\nonumber\\
C_{V^c}&=&C_{V^c0}+\frac{A_2^2(T_{eff}^c,P_{eff}^c)}{2B}T_{eff}^c,\nonumber\\
%\beta_c&=&\beta_{c0}+\frac{A_1(T_{eff}^c,P_{eff}^c)A_2(T_{eff}^c,P_{eff}^c)}{2V^cB},\nonumber\\
\kappa_{T_{eff}^c}&=&\kappa_{T_{eff}^c0}+\frac{A_2^2(T_{eff}^c,P_{eff}^c)}{2V^cB},\nonumber
\end{eqnarray}
which indicate that they both have a jump or a peak at critical point as well as the ferromagnetic systems. The critical exponents are $\alpha'=\alpha''=0$. Since $\chi^{-1}=\frac{\partial^2G}{\partial\eta^2}=2A+12B\eta^2$ when $P_{eff}=P_{eff}^c$, we have
\begin{eqnarray}
\chi=\left\{
  \begin{array}{ll}
   \frac{1}{2A_1(T_{eff}^c,P_{eff}^c)(T_{eff}-T_{eff}^c)}& ~~~~~~~~~~~\text{for}~~~~~ T_{eff}>T_{eff}^c \\
    \frac{1}{A_1(T_{eff}^c,P_{eff}^c)(T_{eff}^c-T_{eff})} & ~~~~~~~~~~~\text{for}~~~~~ T_{eff}<T_{eff}^c
  \end{array}\right.,\label{phi}
\end{eqnarray}
and the critical exponents are $\gamma=\gamma'=1$. Furthermore, for the RN-dS spacetime there exist the same scale relations
\begin{eqnarray}
\alpha'+2\beta'+\gamma=2,~~~\alpha'+\beta'(\delta+1)=2.
\end{eqnarray}

\section{Thermodynamic geometry of RN-dS spacetime}
\label{Six}
As shown in the last section, the critical exponents are all independent with $A$ and $B$, as well as an ordinary thermodynamic system. The reason is that the fluctuation of the order parameter $\eta$ near the critical point is neglected. In Ref. \cite{Ruppeiner2008}, the authors investigated the phase transition structure of black holes through the singularity of the spacetime scalar curvature $R$. The scalar curvature is divergent at the critical point of AdS black hole phase transition as fluid systems do. This property can be used to investigate the information contained in RN-dS spacetime phase transition from the thermodynamic geometric perspective. The sign of $R$ can be used as a diagnostic of the interaction between two microscopic constituents (or molecules) of this system. A positive scalar curvature indicates a repulsive interaction, whereas a negative one stands for an attractive interaction. In this part, we will investigate the scalar curvature to reveal the microstructure of this system.

\begin{figure}[htp]
\includegraphics[width=0.4\textwidth]{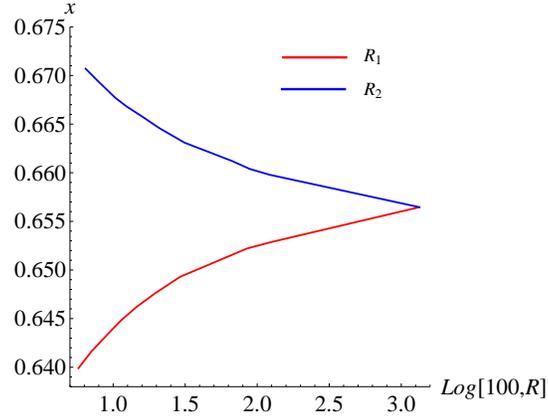}
\caption{The scalar curvature of two phases in the coexistent states.}\label{R-x}
\end{figure}

For simplicity, we adopt the fluctuation coordinates $T_{eff}$ and $V$, the metric is two-dimensional and diagonal
\begin{eqnarray}
g_{\mu\nu}=\frac{1}{T_{eff}}\left(
                              \begin{array}{cc}
                                \frac{\partial S}{\partial T_{eff}}\big|_V & 0 \\
                                0 & \frac{\partial P_{eff}}{\partial V}\big|_{T_{eff}} \\
                              \end{array}
                            \right).
\end{eqnarray}
Since the heat capacity $C_V=T_{eff}\frac{\partial S}{\partial T_{eff}}\big|_V$, the line element can be expressed as
\begin{eqnarray}
dl^2=\frac{C_V}{T_{eff}^2}dT_{eff}^2+\frac{(\partial_V P_{eff})|_{T_{eff}}}{T_{eff}}dV^2,
\end{eqnarray}
the scalar curvature is easily worked out
\begin{equation}
\begin{split}
R=&\frac{1}{2C^2_V(\partial_V P_{eff})}\left\{T_{eff}(\partial_V P_{eff})\left[(\partial_VC_V)^2+\partial_{T_{eff}}C_V(\partial_V P_{eff}-T_{eff}\partial_{T_{eff},V}P_{eff})\right]\right.\\
&\left.+C_V\left[(\partial_V P_{eff})^2+T_{eff}(\partial_VC_V(\partial_{V,V}P_{eff})-T_{eff}(\partial_{T_{eff},V}P_{eff})^2)\right.\right.\\
&\left.\left.+2T_{eff}\partial_V P_{eff}(T_{eff}(\partial_{T_{eff},T_{eff},V}P_{eff})-\partial_{V,V}C_V)\right]\right\}.\label{R}
\end{split}
\end{equation}
Combining eqs. (\ref{V}), (\ref{T}), (\ref{P}), (\ref{Cv}), and (\ref{R}), the behaviours of the scalar curvature for two different phases with the horizons' ratio $x$ is displayed in Fig. \ref{R-x}.
From Fig. \ref{R-x}, we find that the values of scalar curvature for two-coexistent phases are both positive, which means the average interaction of the system in two different coexistent phases is repulsive. Since $R_2\geq R_1$, the interaction of the phase $\phi_{+2}$ is bigger than the one with the phase $\phi_{+1}$, until they are equal to each other at the critical point.

\section{Conclusion}
\label{Seven}
In this work, when considering the gravity effect and the interplay between the black hole and cosmological horizons of the four-dimensional RN-dS spacetime, we presented the effective thermodynamic quantities of RN-dS spacetime in thermodynamic equilibrium from the viewpoint of the electric potential on black hole horizon ($\phi_+$). Based on these effective quantities, the thermodynamical properties of RN-dS spacetime were investigated. This provides a good way to further understand the relationship between the gravity and the thermodynamics of black holes.

First, with a certain set of parameters we gave the condition of the ratio between the black hole and the cosmological horizons which were both survived in the four-dimensional RN-dS spacetime. And the concrete process of exporting the effective thermodynamic quantities and critical point of RN-dS spacetime were also presented. Through the analyse of the heat capacities, the isobaric expansion coefficient, and the isothermal compression coefficient, respectively, we found that there exist a peak behavior both of these quantities which are as the function of $x$. Those results indicate the existence of the second-order phase transition of this system in the certain parameters. Then, the first-order phase transition in the induced phase space was also exhibited by two methods: the Maxwell's equal area law and Gibbs free energy, which is independent on the model parameters. And the two-phase coexistent curve in the $P-T$ plane was presented.

For this system, the Prigogine-Defay (PD) ratio equals to one which is consistent with that in the ordinary thermodynamical systems as well as AdS black holes. In order to describe the  thermal expansion with an isenthalphy process, the properties of the isoenthalpy and inversion curves for this system were also  shown. The isoenthalpic curve is divided into two parts by the corresponding inversion curve: the cooling phenomena with the positive slope and the heating process with the negative slope of the $P-T$ curve. In addition, based on the Landau's continuous theory of phase transition we presented the critical exponents and the scale relations, which are consistent with AdS black holes as well as the ferromagnetic systems. Those similarities among in the RN-dS spacetime, AdS black holes, and the VdW systems imply that the RN-dS spacetime and AdS black holes should be of the internal microstructure as well as the VdW system.

Finally, to probe the microstructure of the RN-dS spacetime we introduced the scalar curvatures of two coexistent phases by the Ruppeiner geometry method that is used in AdS black hole thermodynamics. The results indicate the average interactions of the system in two different coexistent phases are repulsive. Furthermore the interaction of the high-potential phase $\phi_2$ is bigger than the one with the low-potential phase $\phi_1$, until they are equal to each other at the critical point.

\section*{Acknowledgments}
We would like to thank Prof. Zong-Hong Zhu and Li-Chun Zhang for their indispensable discussions and comments. This work was supported by the Natural Science Foundation of China (Grant No. 12075143).

\end{document}